\documentclass[pra,onecolumn,nofootinbib,superscriptaddress]{revtex4-2}

\usepackage{lineno}
\usepackage[table]{xcolor} 
\usepackage{array}
\usepackage{tcolorbox}
\usepackage{booktabs} 
\usepackage{bm}
\usepackage{physics}
\usepackage{bbm}
\usepackage{amsmath}
\usepackage{amssymb}
\usepackage{multirow}

\begin{document}

\title{Quantum Optical Techniques for Biomedical Imaging}

\author{Vahid Salari}
\affiliation{Institute for Quantum Science and Technology, Department of Physics and Astronomy, University of Calgary, Calgary, AB, Canada}

\author{Yingwen Zhang}
\affiliation{Nexus for Quantum Technologies, University of Ottawa, Ottawa ON Canada, K1N6N5}
\affiliation{National Research Council of Canada, 100 Sussex Drive, Ottawa ON Canada, K1A0R6}

\author{Sepideh Ahmadi}
\affiliation{Institute for Quantum Science and Technology, Department of Physics and Astronomy, University of Calgary, Calgary, AB, Canada}

\author{Dilip Paneru}
\affiliation{Nexus for Quantum Technologies, University of Ottawa, Ottawa ON Canada, K1N6N5}
\affiliation{National Research Council of Canada, 100 Sussex Drive, Ottawa ON Canada, K1A0R6}
\affiliation{Dipartimento di Fisica, Universit\`{a} degli Studi di Napoli Federico II, Complesso Universitario di Monte Sant'Angelo, Via Cintia, 80126 Napoli, Italy}

\author{Duncan England}
\affiliation{National Research Council of Canada, 100 Sussex Drive, Ottawa ON Canada, K1A0R6}

\author{Shabir Barzanjeh}
\affiliation{Institute for Quantum Science and Technology, Department of Physics and Astronomy, University of Calgary, Calgary, AB, Canada}

\author{Robert Boyd}
\affiliation{Nexus for Quantum Technologies, University of Ottawa, Ottawa ON Canada, K1N6N5}
\affiliation{Institute of Optics, University of Rochester, Rochester, 14627, NY, USA}

\author{Ebrahim Karimi}
\affiliation{Nexus for Quantum Technologies, University of Ottawa, Ottawa ON Canada, K1N6N5}
\affiliation{National Research Council of Canada, 100 Sussex Drive, Ottawa ON Canada, K1A0R6}
\affiliation{Institute for Quantum Studies, Chapman University, Orange, California 92866, USA}

\author{Christoph Simon}
\affiliation{Institute for Quantum Science and Technology, Department of Physics and Astronomy, University of Calgary, Calgary, AB, Canada}
\affiliation{Hotchkiss Brain Institute, University of Calgary, Calgary, AB}

\author{Daniel Oblak}
\affiliation{Institute for Quantum Science and Technology, Department of Physics and Astronomy, University of Calgary, Calgary, AB, Canada}

\begin{abstract} 
Quantum imaging is emerging as a transformative approach for biomedical applications, applying nonclassical properties of light, such as entanglement, squeezing, and quantum correlations, to overcome fundamental limits of conventional techniques. These methods promise superior spatial resolution, enhanced signal-to-noise ratios, improved phase sensitivity, and reduced radiation dose, for potentially safer and more precise imaging for delicate biological samples. Here, we present an overview of quantum optical biomedical imaging technologies as well as quantum-inspired imaging methods, including quantum optical coherence tomography, quantum optical microscopy, ghost imaging, multi-parameter quantum imaging, and imaging with quantum-grade cameras. We describe the operating principles, biomedical applications, and unique advantages of each approach, along with the specific challenges for their translation into real-life practice. This review aims to guide future research toward advancing quantum imaging from experimental demonstrations to impactful biomedical tools.
\end{abstract}
\maketitle
\section{Introduction}

Biomedical imaging techniques are central to diagnosing and monitoring a wide range of diseases. Before the discovery of X-rays in 1895, the only reliable means of visualizing internal anatomy was invasive surgery, which carried significant trauma and risk for patients~\cite{bib1}. Today, biomedical imaging supports accurate diagnosis, therapeutic guidance, and disease monitoring across a wide range of conditions without surgery and almost non-invasively~\cite{bib2, bib3, bib4, bib5, bib6, bib7, bib8, bib9, bib10, bib11, bib12, bib13, bib14, bib15, bib16}. In a typical imaging procedure, energy from a suitable source is directed toward the body, where it interacts with tissues through absorption, scattering, or transmission. The modified signals are then captured by detectors and processed by computational algorithms to reconstruct images. Optical imaging techniques are convenient, less expensive, and safe techniques for biomedical imaging~\cite{ShiAlfano2017DeepImaging, bib78, bib79}. Fluorescent imaging~\cite{bib80}, photoacoustic imaging~\cite{bib81}, endoscopy~\cite{bib82}, diffuse optical tomography (DOT)~\cite{bib83, bib84}, super-resolution microscopy~\cite{bib86} and optical coherence tomography (OCT)~\cite{bib87, bib88, bib89, bib90, bib91, bib92, bib93} are a few examples among others. Recently, some of these techniques have been further developed by using quantum optical techniques~\cite{MandelWolf1995,WallsMilburn2008}, leading to the development of quantum imaging~\cite{Boyd2003NonlinearOptics, Moreau2019NRP,Genovese2016RealApplicationsQI}. 

Quantum mechanics, developed in the first half of the twentieth century, has significantly enhanced measurement accuracy and sensitivity by exploiting non-classical features, particularly nonclassical correlations~\cite{Giovannetti2011AdvancesQM,Giovannetti2004,Caves1981}. Quantum optical imaging is supposed to overcome the limitations of classical optical imaging, producing images with higher resolution and sensitivity by using quantum properties of light such as quantum entanglement \cite{Brida2010SubShotNoise,Genovese2016RealApplicationsQI,Moreau2019NRP}. These techniques promise improved performance compared to conventional imaging systems, reduced radiation exposure, and perform better in high-noise environments~\cite{Brida2010SubShotNoise,Tan2008PRLQI,Lloyd2008ScienceQI}. Some of these quantum imaging techniques include Interaction-Free (IF) imaging~\cite{ElitzurVaidman1993IFM,Kwiat1995IFM}, quantum ghost imaging (QGI)~\cite{Pittman1995TwoPhotonImaging,ErkmenShapiro2010AOP}, Quantum Illumination (QI)~\cite{Lloyd2008ScienceQI,Tan2008PRLQI, PhysRevLett.114.080503, Gallego, Barzanjehradar}, Quantum Optical Coherence Tomography (QOCT)~\cite{Abouraddy2002QOCT,Hong1987HOM}, and other techniques based on quantum optical sensing methods~\cite{Degen2017RMP,Giovannetti2011AdvancesQM}. Further advantages of quantum imaging are in biological measurements, such as quantum-enhanced particle tracking~\cite{bib117, bib118, bib119, bib120, bib121}, measuring biomagnetic fields~\cite{bib122, bib123, bib124, bib125, bib126}, refractive index sensing of protein solutions~\cite{bib127}, and microrheology~\cite{bib128, bib129}.

\section{Quantum light sources}

\begin{figure}[h!]
  \centering
  \includegraphics[width=0.6\linewidth]{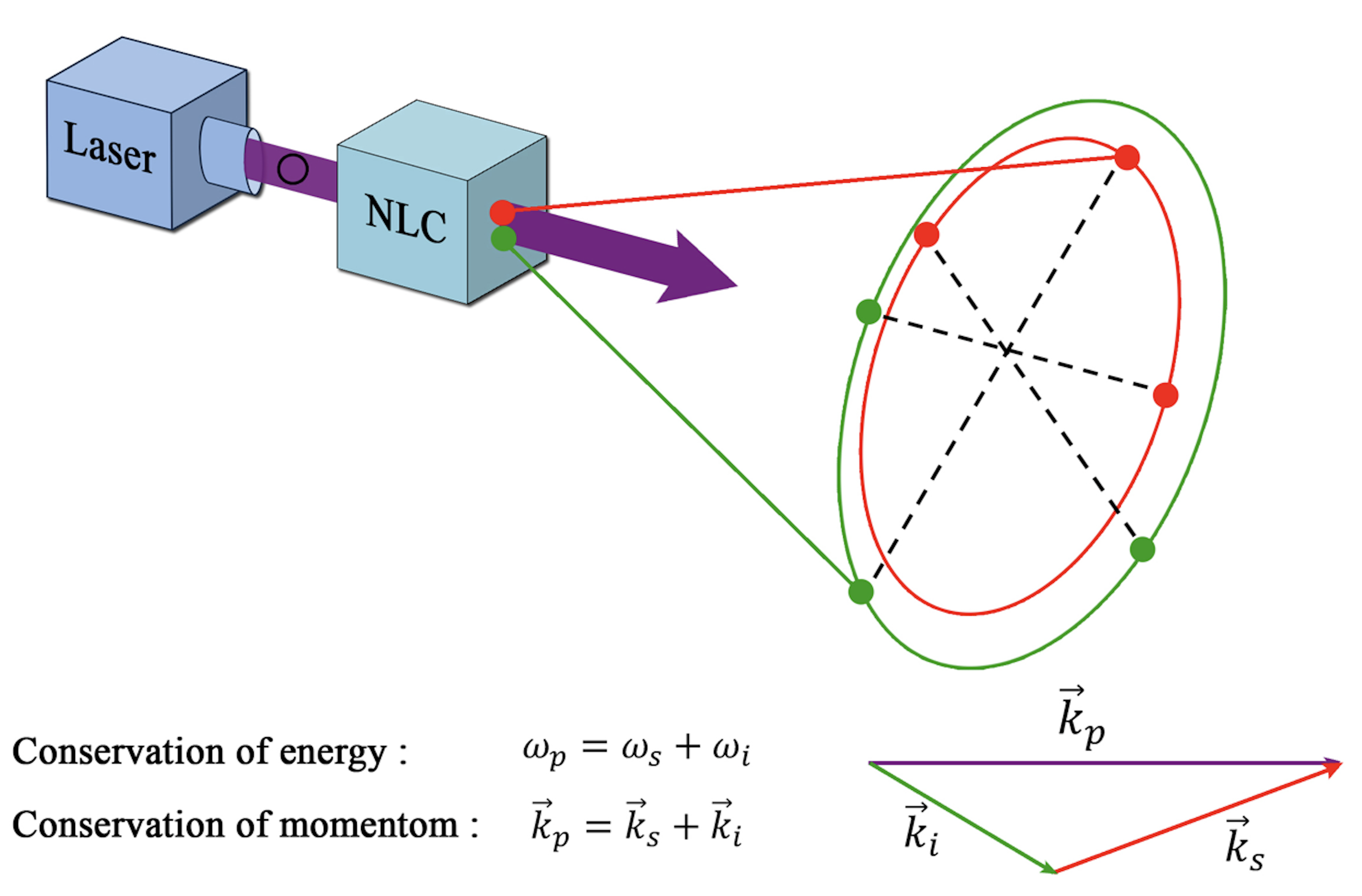}
  \caption{Schematic of a SPDC-based quantum entangled light beams generation, where the conservation of the total energy and momentum in the SPDC process are held}
\label{f12}
\end{figure}

Quantum imaging typically employs pairs of entangled photons. In one configuration, both of these photons interact with the target before being separately measured by their respective detectors. Alternatively, the photons are separated and one photon of the pair, known as the signal photon, interacts with the object and is then collected by a detector. The idler photon, which is entangled with the signal photon, is used as a reference, and is collected by a separate detector (see Fig.~\ref{f12}(a)). Typically, to obtain an image of the object, the time and position correlation between the two photons are measured. However, it is also possible to utilize the correlations in additional degrees of freedom, such as polarization, spectrum and momentum, to gain further information about the object.

The advantages of quantum imaging through entanglement over classical imaging techniques are gained by analyzing the quantum correlations between the entangled photon pairs. Under certain conditions, these correlations make it possible to see details smaller than the diffraction limit~\cite{Boto2000,Giovannetti2004}, reduce noise below the shot-noise level~\cite{Xiao1987,GerryKnight2005}, and filter out unwanted background light~\cite{Pittman1995,Shih2011}.
The sub shot-noise and background filtering ability allow quantum imaging methods to work with fewer photons, thereby exposing the object to less radiation~\cite{bib115, bib116}. This makes them especially useful for studying fragile samples that could be damaged by light or that need to stay at very low temperatures, where weak illumination is a must to avoid heating.

Below, we list some of the most commonly used quantum light sources and quantum states employed for quantum imaging.\\

\textit{Spontaneous parametric down-conversion:} One of the most common processes for the creation of entangled photon pairs is through the process of spontaneous parametric down-conversion (SPDC)~\cite{bibz1,bibz2}. In this process, a laser (light beam) at frequency $\omega_{P}$ is used to pump a nonlinear crystal with a large second-order nonlinearity $\chi^{(2)}$, such as $\beta$-barium borate (BBO), lithium niobate (LiNbO$_3$) or potassium titanyl phosphate (KTiOPO$_4$) crystals. In SPDC, a pump photon at frequency $\omega_p$ is converted into a pair of lower-frequency photons, called signal and idler ($\omega_s, \omega_i$), mediated by the second-order nonlinear polarization $P^{(2)}(t) = \epsilon_0 \chi^{(2)} E^2(t)$ where $P^{(2)}(t)$ is the second-order nonlinear polarization, $\epsilon_0$ is the vacuum permittivity, $\chi^{(2)}$ is the second-order nonlinear susceptibility of the medium, and $E(t)$ is the total electric field, including the pump, signal, and idler components. The nonlinear susceptibility $\chi^{(2)}$ couples the pump, signal, and idler fields, enabling energy conservation and phase-matching conditions for efficient generation.
As shown in Fig.~\ref{f12}, energy conservation requires $\omega_p = \omega_s + \omega_i$ and momentum conservation requires $k_{p} =  k_{s} + k_{i}$, where $\omega_{J}$ and $k_{J}$, for $J=s,i,p$ are the frequencies and propagation vectors for the signal, idler, and pump photons, respectively. In fact, the signal and idler beams are spatially and temporally correlated as they are created at the same time and position through the same pump photon.\\ 

\textit{Spontaneous four-wave mixing:} Another commonly used technique for generating temporally and spectrally entangled photons is through the process of spontaneous four-wave mixing (SFWM), whereby two photons from a pump laser are converted to a pair of entangled photons through a third-order nonlinear process. SFWM typically requires significantly longer nonlinear media than SPDC. Optical fibers are commonly used as the nonlinear material for SFWM, and as a result, position and momentum correlation are lost through the propagation inside the fiber, and only temporal and spectral entanglement remains. SFWM is generated from the nonlinear polarization $P^{(3)}(t) = \epsilon_0 \chi^{(3)} E^3(t)$, where $P^{(3)}(t)$ is the third-order nonlinear polarization, $\epsilon_0$ is the vacuum permittivity, $\chi^{(3)}$ is the third-order nonlinear susceptibility of the medium, and $E(t)$ is the total electric field, which here includes the pump, signal, and idler contributions. The two pump photons at frequency $\omega_p$ are annihilated to create a signal and idler photon at $\omega_s$ and $\omega_i$. The process obeys strict energy and momentum conservation conditions, $2\omega_p = \omega_s + \omega_i$ and $2k_p = k_s + k_i + \Delta k$~\cite{Sharping2001,Lin2007}. Because position–momentum correlations are lost within the fiber, SFWM is more often employed in quantum communication than in quantum imaging.~\cite{Fan2005, Afsharnia2024, Wang2024, GarayPalmett2007, Ma2017}. \\

\textit{N00N states:} A N00N state is a state of entanglement between an N-photon state and a vacuum state, in quantum optics, typically written as~\cite{Lee2002, Walther2004, Afek2010, Israel2014, Dowling2008}.
\begin{equation}
    \ket{N00N} = \frac{1}{\sqrt{2}}\left(\ket{N}_a\ket{0}_b + e^{i\phi}\ket{0}_a\ket{N}_b \right),
\end{equation}
where $\ket{N}_j$ is a Fock state of N photons in mode j, $\ket{0}_j$ is the vacuum state in mode j, and $\phi$ is a phase.

Experimentally, $N=2$ NOON states are typically generated using SPDC as a photon source, combined with linear optical elements such as beam splitters, phase shifters, and post-selection based on photon-number-resolving detectors. Two-photon N00N states can be generated directly through Hong-Ou-Mandel interference. N00N states are useful in improving the phase sensitivity of imaging systems and exceed diffraction-limited spatial resolution, e.g. quantum lithography that uses multiphoton entangled states, i.e. most famously N00N states, to achieve patterning with resolution beyond the classical diffraction limit~\cite{bib185}. 

 Experiments have demonstrated three-photon N00N states ($N=3$) through heralded generation via SPDC, for super-resolved interference fringes~\cite{Israel2014,Kim2009, Boyd2005QuantumLithography}. In addition, N00N states up to four photons ($N=4$) have been characterized via quantum state tomography~\cite{Israel2011}. \\

\textit{Squeezed States:} Squeezed light is based on correlations in quadrature variables that go beyond the shot-noise limit~\cite{Braunstein2005,Kolobov2007}. A two-mode squeezed vacuum state is typically described as
\begin{equation}
|\psi\rangle = \exp\!\big[r \,(a b - a^\dagger b^\dagger)\big] |0,0\rangle,
\end{equation}
where $r$ is the squeezing parameter, and $a, b$ ($a^\dagger, b^\dagger$) are the annihilation (creation) operators of the two modes, and the strength of entanglement is specified with the magnitude of the squeezing parameter, $r$. In quantum imaging, such entanglement supports noise reduction below the standard quantum limit, and improves contrast or phase sensitivity, providing the visualization of weakly emitting or low-contrast samples with higher fidelity than classical light sources~\cite{Kolobov2007}.

Experimentally, squeezed states are typically generated via nonlinear optical processes such as parametric down-conversion in $\chi^{(2)}$ crystals or four-wave mixing in $\chi^{(3)}$ media, where quantum fluctuations in one quadrature are suppressed below the shot-noise limit~\cite{WallsMilburn2008}. For detection, balanced homodyne or heterodyne setups are widely used, while imaging applications often apply highly sensitive CCD/EMCCD or sCMOS cameras combined with spatially resolved homodyne detection to capture quantum correlations across the image plane~\cite{Treps2003}.

\section{Quantum optical imaging}

In the following, we will focus on a few important techniques, including imaging using quantum states of light in quantum optical coherence tomography, quantum optical microscopy, quantum ghost imaging, and multi-parameter quantum imaging, as well as techniques using quantum sensors for imaging biological ultraweak photon emission.

\subsection{Quantum optical coherence tomography}
Conventional optical coherence tomography (OCT) is one of the most useful medical diagnostic methods. It is based on a Mach-Zehnder interferometer using low-coherence light to provide sub-surface images (B-scans) of the retina, skin, or coronary artery~\cite{bib87, bib88, bib89, bib90, bib91, bib92}. However, the axial resolution of OCT is limited by the coherence length of the light source and affected by material dispersion, which reduces image quality and depth accuracy. The quantum version of OCT, named quantum OCT (QOCT), utilizes entangled photons and Hong-Ou-Mandel (HOM) interferometry. It provides greater resilience to dispersion and provides double the axial resolution of the image~\cite{bib192, bib193, bib194, bib195, bib196}. In QOCT, entangled photon pairs generated via SPDC are directed into a HOM interferometer, with one photon probing the sample and the other traveling along the reference arm. The central observable is the coincidence rate $R_c(\tau)$ as a function of the delay $\tau$ 
between the two arms:
\begin{equation}
R_c(\tau) \propto \int d\Omega \, \big| \phi(\Omega) \big|^2 
\left[ 1 - \cos\left( \Omega \tau + \Delta \phi(\Omega) \right) \right],
\end{equation}
where $\phi(\Omega)$ is the joint spectral amplitude of the photon pair (generated through the SPDC process) and $\Delta \phi(\Omega)$ encodes the phase accumulated in the sample arm.  

In the absence of sample dispersion, the coincidence rate reduces to the well-known Hong–Ou–Mandel dip:
\begin{equation}
R_c(\tau) \propto 1 - \exp\left(- \frac{\tau^2}{2 \tau_c^2} \right),
\end{equation}
where $\tau_c$ is the biphoton coherence time. The full-width at half-maximum (FWHM) of this dip determines the axial resolution, which is twice that of conventional OCT for the same source bandwidth~\cite{Abouraddy2002,Nasr2003}. 

The sample reflectivity profile $r(z)$ can be retrieved by scanning the delay $\tau$ and recording $R_c(\tau)$:
\begin{equation}
R_c(\tau) \propto \int dz \, |r(z)|^2 \, h(\tau - 2z/c),
\end{equation}
where $h(\tau)$ is the biphoton coherence envelope and $c$ is the speed of light. The integration in Eq.~(5) does not eliminate $r(z)$ by averaging, it converts the complex reflectivity into its intensity contribution $|r(z)|^{2}$, which is convolved with the biphoton coherence function $h(\tau)$. Consequently, the depth-dependent reflectivity profile is preserved (up to the system's axial resolution).
This expression assumes a real reflectivity profile. In the full treatment, however, the sample is described by a complex, frequency-dependent reflectivity \(r(z,\Omega)\). When frequency-entangled photon pairs traverse the two arms of the interferometer, the dispersive phase picked up by one photon is correlated with that of its twin. Under symmetric phase-matching, the even-order dispersion terms in the sample’s phase expansion cancel in the coincidence rate, leaving only odd-order contributions (see Refs.~\cite{Abouraddy2002,Nasr2003} for a full derivation).

\begin{figure}[h]%
\centering
\includegraphics[width=1\linewidth]{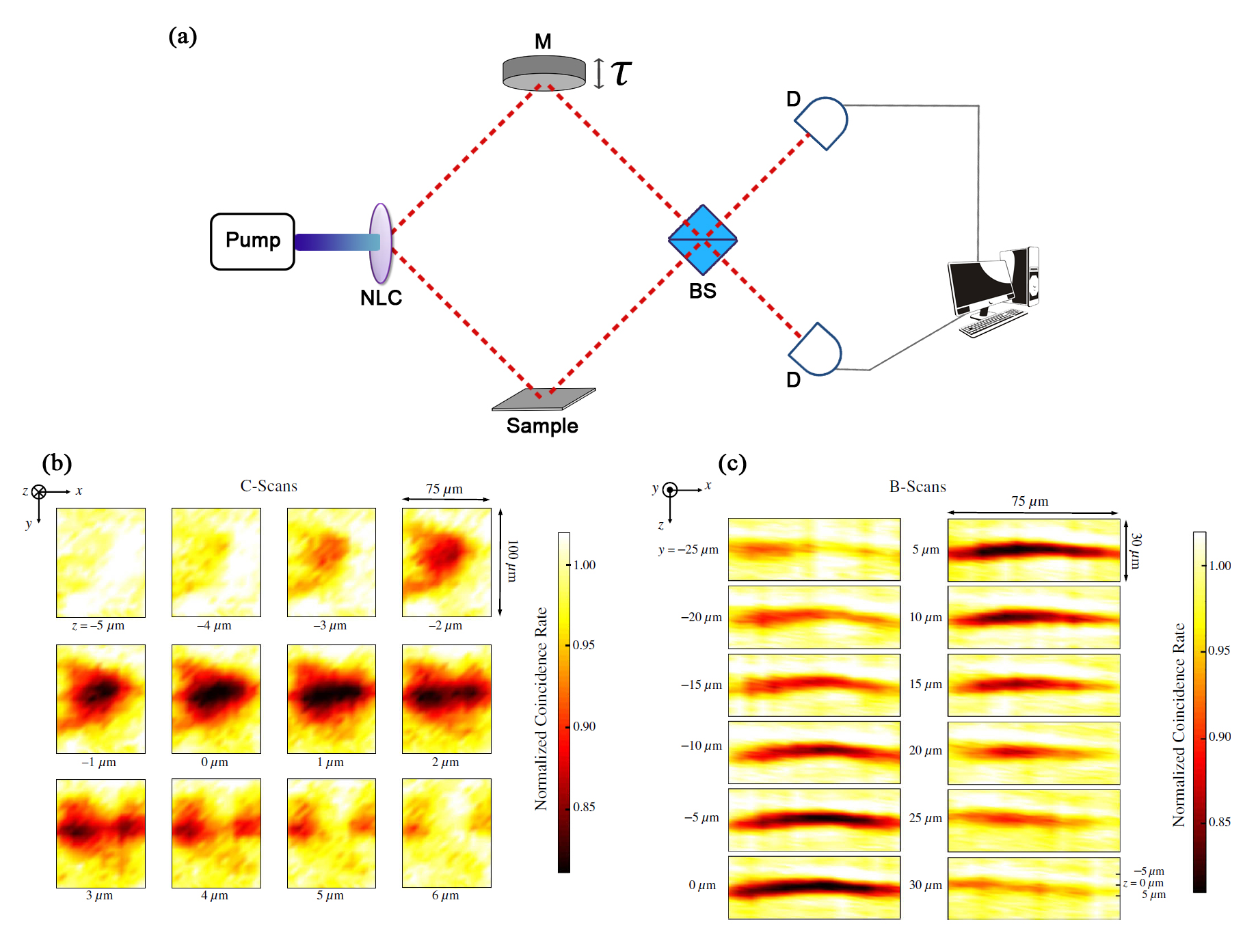}
\caption{(a) A QOCT setup, where two-dimensional transverse (xy) images (C-scans) of an onion-skin sample (coated with gold nanoparticles) acquired at various axial depths (z). Two-dimensional axial (xz) images (B-scans) of the onion-skin are recorded at different transverse positions (y). A pronounced response, corresponding to a reflecting surface, is evidenced by a reduction in the measured coincidence rate. (b) C-scan of onion skin sample in different depths, (c) B-scan of onion skin~\cite{bib197}.}\label{f17}
\end{figure}

A simple schematic of the QOCT setup is shown in Fig.~\ref{f17} for biological imaging, here on onion skin, which is coated with gold nanoparticles~\cite{bib197}. One photon from an entangled pair illuminates the target with the partner reflected from a mirror with a controllable time delay $\tau$. The B-scan is obtained by measuring the coincidence rate against the optical path length difference between photons via a spatially scanned single-photon detector. A B-scan contains information about the number of sample layers and the axial distances between them. By scanning the target transversely on the x–y grid for each axial position z in the B-scan, the 2D image as a C-scan is produced. Subsequently, the 3D image is reconstructed by the stacking of multiple B-scans of one per axial position~\cite{bib198}. Also, QOCT may probe multi-layered samples with superior precision compared to the conventional counterpart~\cite{UofC}.

QOCT has some advantages over conventional OCT, including enhanced resolution, improved signal-to-noise ratio, reduced multiple scattering effects, and superior phase sensitivity and contrast, better penetration depth, and reduced risk of photodamage. However, QOCT typically suffers from low photon flux, limited acquisition speed, and appearance of artifacts \cite{ 2503.06772}, which restricts its practical imaging depth and resolution. Moreover, the complexity of entangled-photon sources and detection schemes increases system cost and reduces robustness compared to the relatively mature and high-speed conventional OCT systems.

\subsection{Quantum sub shot-noise imaging}
The stochastic fluctuation in photon arrival times at the detector, due to the particle nature of light, introduces shot noise. This noise affects the sensitivity, speed, and resolution of conventional microscopy. Sensitivity is constrained by the shot noise limit, where the photon number variance scales as $(\Delta N)^2 = \langle N \rangle$, with $\langle N \rangle$ denoting the mean photon number. Shot noise imposes a phase sensitivity of $\Delta \phi_{\mathrm{SNL}} \sim 1/\sqrt{\langle N \rangle}$, meaning that improvements in resolution or contrast require higher photon flux, which risks photodamage in biological samples ~\cite{bib199, bib200, bib201}. 

Quantum resources, such as squeezed states and N00N states, enable sensitivities beyond the shot-noise limit, reaching toward the much lower Heisenberg limit 
$\Delta \phi_{\mathrm{HL}} \sim 1/\langle N \rangle$~\cite{Caves1981, Giovannetti2004,Taylor2013}. Hence, sub shot-noise quantum imaging applies the properties of quantum states of light, such as correlations between entangled photons and squeezed light, to surpass the classical shot noise limit, achieving sensitivities that are unattainable with conventional light sources. Fluctuations in the photon number, i.e. shot noise, of SPDC sources are spatially and temporally correlated due to the entanglement between the photons. Exploiting this behavior by performing spatial and temporal correlation measurements facilitates reducing these random fluctuations below the shot-noise limit~\cite{bib225, bib227, bib211, bib212}. In contrast, squeezed light involves the manipulation of the quantum state of light to reduce noise in one of its properties (e.g., amplitude or phase) to below the shot noise at the expense of increased noise (anti-squeezing) in the conjugate property~\cite{bib214}. This advantage underpins quantum-enhanced microscopy, where reduced noise and superior phase estimation allow imaging with fewer photons, preserving fragile biological structures and features.

\subsection{Fluorescence and multiphoton microscopy with entangled light} 

Fluorescence microscopy and confocal laser scanning microscopy are important tools in biomedical imaging that use fluorophores. These fluorophores emit photons one at a time and exhibit quantum properties such as sub-Poissonian photon statistics and quantum correlations, which are typically ignored in classical measurements. By recording coincidence detections at each pixel, it is possible to differentiate the emitted fields from their close neighbours, thereby enhancing imaging resolution~\cite{bib215, bib216, bib217, bib218, bib219}. 

\begin{figure}[h]
\centering
\includegraphics[width=1\linewidth]{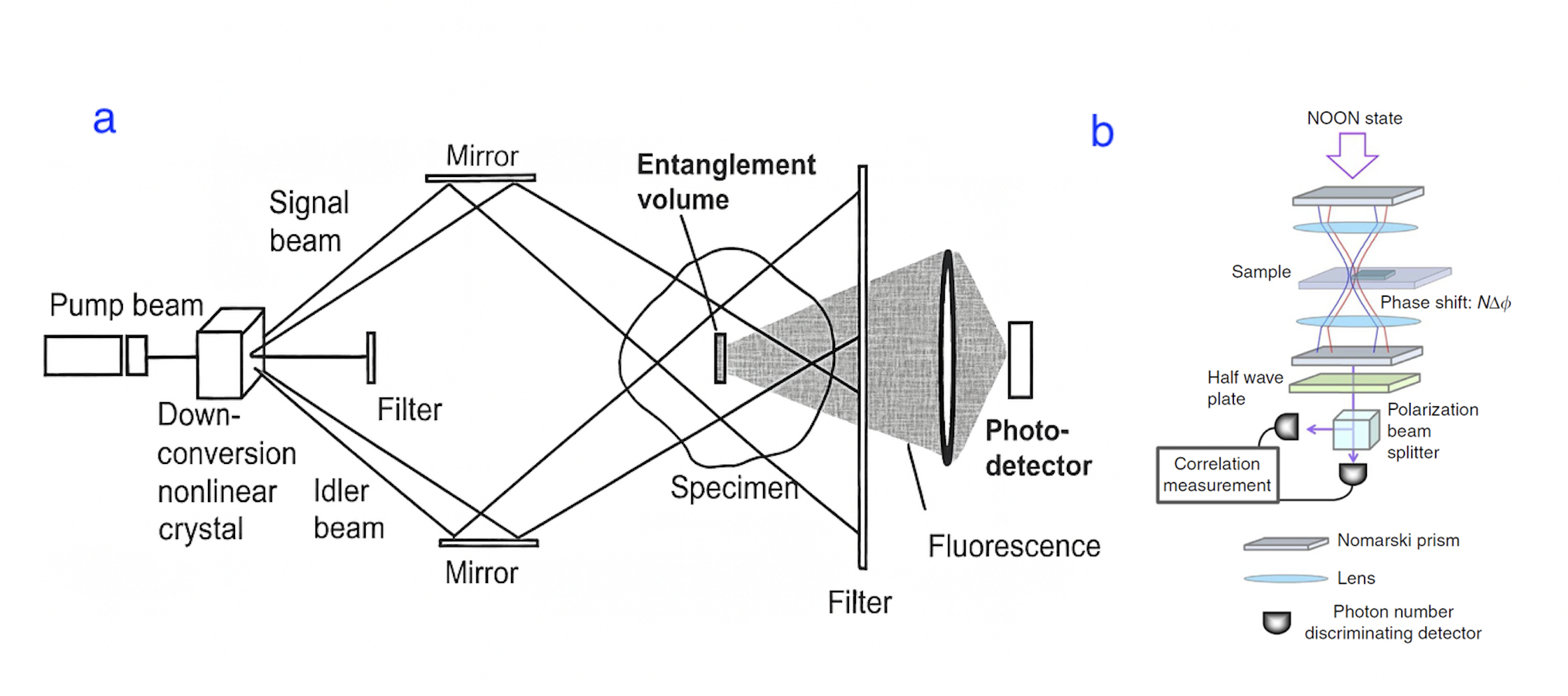}
\caption{(a) A scheme of entangled microscopy in a multi-photon fluorescence microscope where two entangled photons are directed at fluorophores, resulting in photon-pair absorption~\cite{bib202}. Multi-photon absorption by fluorophores reduces the number of photons needed to illuminate the sample with a fixed SNR, minimizing potential damage. (b) Quantum measuring using N00N states, where two correlated light beams illuminate slightly different sections of a sample~\cite{bib208}.}
\label{f18}
\end{figure}

In addition, multi-photon absorption by fluorophores holds significant advantages over single-photon microscopy, such as reducing the number of photons needed to illuminate the sample, as fluorescence occurs only at the focal spot, thereby minimizing potential damage~\cite{bib114}. Figure~\ref{f18}(a) illustrates a multi-photon fluorescence microscope where entangled photons are directed at fluorophores, resulting in photon-pair absorption. Multiphoton fluorescence microscopy relies on nonlinear two-photon absorption, where the excitation rate with classical light scales quadratically with intensity, $R_{\mathrm{cl}} \propto I^2$. This can be expressed in terms of the classical two-photon absorption cross-section $\sigma^{(2)}_{\mathrm{cl}}$ as $R_{\mathrm{cl}} = \sigma^{(2)}_{\mathrm{cl}} I^2$. By contrast, entangled photon pairs, in principle, exhibit a linear scaling, with the corresponding cross-section $\sigma^{(2)}_{\mathrm{ent}}$, such that $R_{\mathrm{ent}} = \sigma^{(2)}_{\mathrm{ent}} I$.

This linear dependence allows efficient multiphoton excitation at much lower photon flux~\cite{Fei1997,Lee2006,Varnavski2020}. The quantum enhancement reduces photodamage while maintaining resolution, apparently making entangled-photon microscopy useful for noninvasive imaging of photosensitive biomolecules and living tissues. However, this claim still remains controversial. Recently, Landes et. al \cite{Landes2024EntangledTPA} systematically examined fluorescence-detected two-photon absorption with time-frequency-entangled photon pairs from low- to high-gain regimes. In the low-flux regime, where quantum enhancement is predicted, they found the molecular fluorescence signal to be below the detection threshold, showing no clear advantage over classical excitation. At higher fluxes they observed fluorescence but the conditions approach classical behavior, blurring any quantum benefit. Their results challenge the claims that entangled two-photon absorption can reduce photodamage  while retaining resolution. Consequently, advantages in  practical bio-imaging remain unverified.

Another approach involves the use of quantum correlations in differential interference microscopy, which improves sensitivity in phase measurements and facilitates images to be acquired with a higher SNR. In this technique, as shown in Fig.~\ref{f18}(b), two correlated light beams illuminate slightly different sections of a sample. The differences between these sections are measured by analyzing the relative phases, which helps reconstruct a full image through raster scanning.

\begin{table}[t]
\renewcommand{\arraystretch}{1.3}
\centering
\caption{Representative benchmarks for fluorescence microscopy. Values are setup-dependent and are not restricted to these numbers; here, they are based on \cite{Denk1990,Taylor2013,Lee2006,Varnavski2020}.}
\label{tab:QFM}
\begin{tabular}{|c|c|c|}
\hline
\textbf{Metric} & \textbf{Conventional microscopy} & \textbf{Quantum-enhanced} \\
\hline\hline
Photon flux / Power & average power 10--100 mW (MHz pulsed lasers) & $10^{5}$--$10^{7}$ pairs/s ($\sim$nW) \\
Frame rate & 1--30 fps & $\ll$1 fps (lab demos) \\
Sensitivity & Shot-noise limited (SNL) & Sub-SNL ($\sim$1–3\,dB); linear two-photon absorption \\
\hline
\end{tabular}
\end{table}

Table ~\ref{tab:QFM} is a comparison between classical and quantum florescence microscopy for three metrics. However, despite the predicted advantages of quantum optical microscopy, it faces challenges such as the complexity of generating and maintaining nonclassical light states with sufficient stability and brightness for biological samples. Moreover, these methods require highly efficient detectors, precise synchronization, and advanced correlation measurements, which can be challenging to scale and integrate into practical biomedical imaging systems in the near term.

\subsection{Quantum ghost imaging}
Quantum ghost imaging (GI) is a technique based on the quantum correlation between entangled photons and has garnered the attention from researchers since the 1990's~\cite{bibz3,bibz4,bib223}. In quantum ghost imaging,  the signal and idler photons, generated through SPDC, are split into two paths. The signal photon is used to interrogate an object and is collected by a single pixel bucket detector with only timing resolution. The idler photon, without interacting with the object, is detected by a multi-pixel detector such as an intensified CCD (ICCD) camera. By triggering the camera shutter with the bucket detector so the photon pairs are detected in coincidence, an image of the object will be formed on the camera through the idler photons as a result of the inherent spatial correlation between the photon pair even though the idler photons never interacted with the object, hence the name ``ghost imaging''~\cite{bib115, bib224,bib226,bib228}. Theoretically \cite{Shapiro2012PhysicsGhostImaging}, this can be described as follows:

In the low-gain regime, the position entanglement of two SPDC photons can be written as
\begin{equation}
    |\Psi\rangle = \int\int  \psi(\bm{r}_s,\bm{r}_i) |\bm{r}_s\rangle |\bm{r}_i\rangle\,d^2r_s d^2r_i,
\end{equation}
where $\bm{r}_s,\bm{r}_i$ are the signal and idler's photons transverse positions, and $\psi(\bm{r}_s,\bm{r}_i)$ is the biphoton correlation function which can be written, to good approximation, as
\begin{equation}
    \psi({\bm{r}}_s,{\bm{r}}_i) \propto \exp\left(\frac{-|\bm{r}_s-\bm{r}_i|^2}{2\delta_r^2}\right)\exp\left(\frac{-|\bm{r}_s+\bm{r}_i|^2}{2w_p^2}\right).
\end{equation}
Here, $\delta_r$ and $w_p$ are the position correlation and the pump beam widths, respectively. When illuminating a target $T(\bm{r})$ with the signal photon, then collecting with a bucket detector. The coincidence pattern seen by the idler photon is given by
\begin{equation}
    C(\bm{r}_i) \propto \Big|\int  T(\bm{r}_s)\psi({\bm{r}}_s,{\bm{r}}_i)\,d^2r_s \Big|^2.
    \label{Cimag}
\end{equation}
If assuming a plane wave where $w_p \gg \delta_r$, the coincidence pattern given by Eq.~\eqref{Cimag} becomes
\begin{equation}
    C(\bm{r}_i) \propto \Big|\int T(\bm{r}_s)\exp\left(\frac{-|\bm{r}_s-\bm{r}_i|^2}{2\delta_r^2}\right)\,d^2r_s\Big|^2,
\end{equation}
which is the convolution between the target profile and a Gaussian of width $\delta_r$.

\begin{figure}[h]%
\centering
\includegraphics[width=1\linewidth]{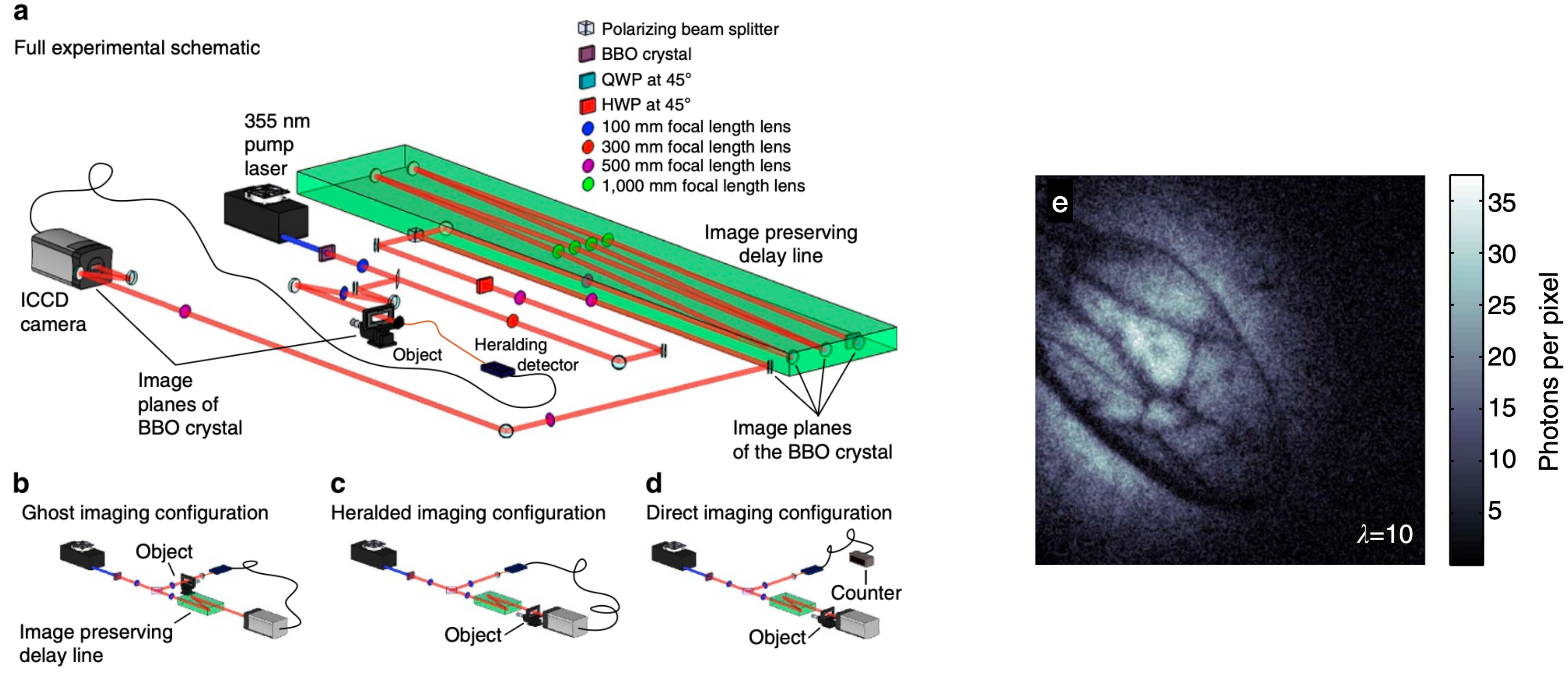}
\caption{Quantum imaging with a very low intensity light of SPDC (a) Full imaging experiment setup where a 355~nm laser pumps a BBO crystal, generating collinear down-converted photon pairs at 710~nm. The crystal’s output facet is imaged onto both the microscope slide (holding the sample) and then onto the ICCD camera. An image-preserving delay line is included to compensate for electronic delays in the triggering system~\cite{bib115}. (b) Ghost imaging - The object (sample) is positioned in the heralding arm, and the camera is triggered by photon detections at the heralding detector. (c) Heralded imaging – The object is placed in the camera arm, while heralding detector events still trigger the camera. (d) Direct imaging – The object remains in the camera arm, but the camera is triggered internally at a rate matched to the single-photon counts measured by the counter in the heralding arm. (e) Ghost image (reconstructed) from non-local imaging of setup (b) from a light-sensitive biological sample, from a weakly absorbing wasp wing with the scale bar of 400~$\mu$m \cite{bib115}.}
\label{f20}
\end{figure}

Quantum GI can image through scattering media~\cite{Dixon2011,Chan2011} and can also operate in a nondegenerate regime where the photon pairs have different wavelengths, which can optimize detection by assigning idler photons to the camera’s sensitive range while tailoring signal photons to the sample’s spectral response~\cite{Aspden2015}. Quantum GI has also inspired classical variants (see sec. 4.1), including thermal-light~\cite{bibz11,bibz12,bibz15,bibz16} and computational ghost imaging~\cite{Shapiro2008,Sun2013}. The configuration of GI and an example of imaging from natural tissue are illustrated in Fig. \ref{f20}. 

Generally, both classical and quantum GI arise from the photon flux-density cross-correlation between the fields received by the bucket and reference detectors. Compared to classical GI, quantum GI presents higher contrast and visibility, and also a wider field-of-view~\cite{bibz22,bib226,bibz24}. Additionally, a significant featureless background appears in classical GI images, something not present in quantum GI~\cite{bibz17, bibz21}. This results in a lower SNR in classical GI compared to quantum GI at equal photon numbers, giving quantum GI an advantage at low light levels~\cite{bibz25,bibz26}. It should also be noted that quantum GI does not provide a  resolution advantage compared to classical GI~\cite{bibz27}. Quantum GI currently faces challenges in producing bright, stable entangled photon sources and requires highly efficient detectors and synchronization, still limiting its practicality for real-world imaging.

\textit{Heralded imaging} is a similar imaging technique to GI based on correlated photon pairs generated through SPDC. In heralded imaging, the camera directly collects photons that pass through (or reflect from) the object, but only records them when a heralding photon is detected~\cite{bib115}. This heralding ensures that only genuine photon-pair events are registered, which improves the signal-to-noise ratio and reduces background (see Fig.~\ref{f20}a and c). In ghost imaging, the object’s information is inferred from correlations, i.e. image formation without direct photon–object interaction (Fig.~\ref{f20}a and b).

\textit{Interaction-free ghost imaging (IFGI)} combines the principles of ghost imaging with interaction-free measurement, a technique using the property of single photon interference to detect the presence of an object without or with reduced photon interaction ~\cite{bibz51}. IFGI can be used to further reduce the photon dosage in quantum GI. 

Also, a high-contrast IFGI can approach background-free performance at very low flux; in practice, dark counts and stray light set a floor~\cite{Ahmadi2023}, making it well-suited for imaging delicate or photosensitive biological tissues. By minimizing photon absorption while maintaining image fidelity, IFGI presents a promising route for noninvasive imaging of sensitive biological samples such as neural tissue or the retina. 

\subsection{Quantum imaging through induced coherence:} 
\begin{figure}
    \centering
    \includegraphics[width=1\linewidth, trim=1cm 4cm 1cm 1cm, clip]{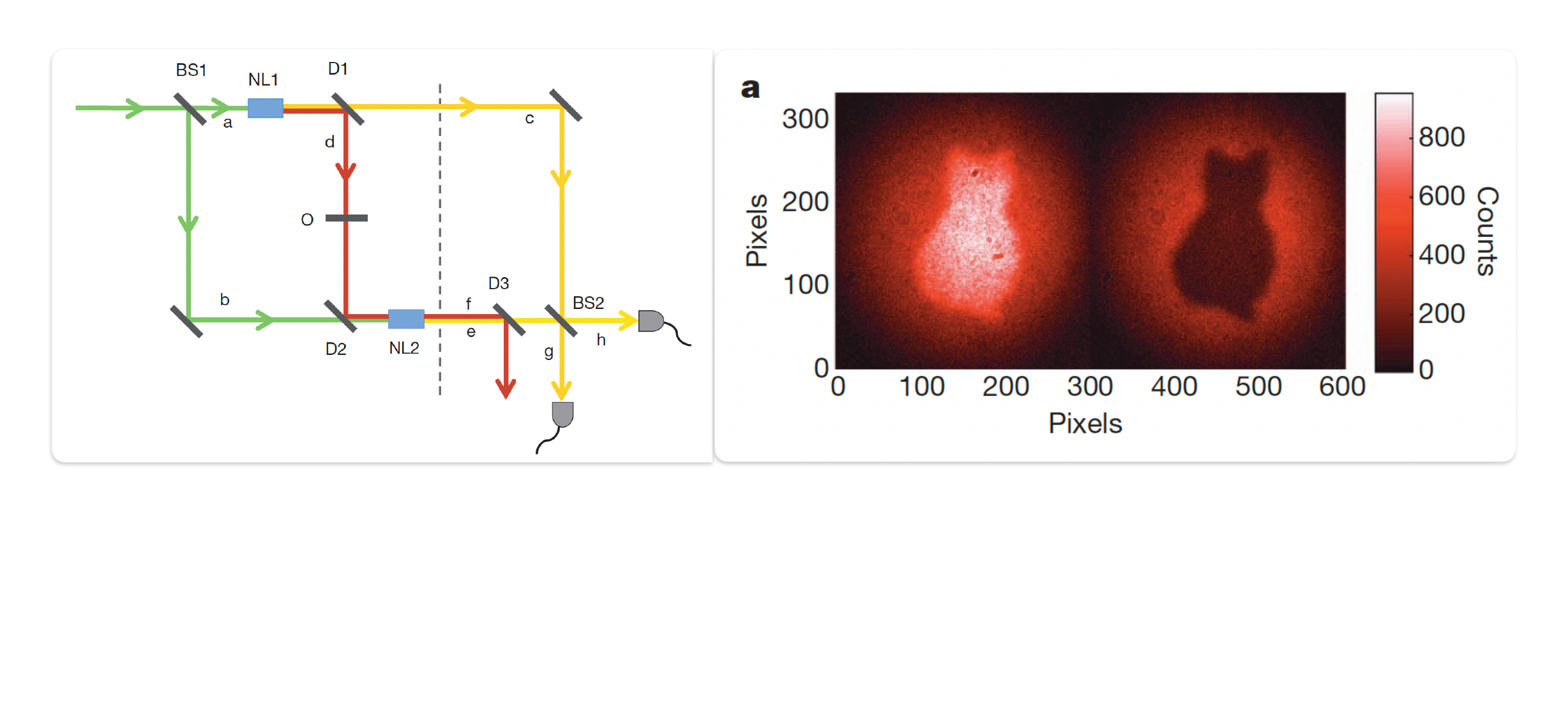}
    \caption{Quantum imaging with undetected photons. Left) Green laser light is divided by the beam splitter BS1 into two paths, modes $a$ and $b$. 
Path $a$ pumps the nonlinear crystal NL1, where collinear SPDC may 
generate a pair of photons at different wavelengths, referred to as the signal (yellow) and idler (red). After traversing the object O, the idler photon is reflected by the dichroic 
mirror D2 so that it becomes indistinguishable from the idler produced in NL2. As a result, 
the final emerging idler state $\lvert f \rangle_{i}$ carries no information about which crystal generated 
the photon pair. Consequently, the signal states $\lvert c \rangle_{s}$ and $\lvert e \rangle_{s}$ interfere 
at the beam splitter BS2, and the output signal beams $\lvert g \rangle_{s}$ and $\lvert h \rangle_{s}$ 
reveal the idler’s transmission properties through the object O. Right) Inside the interferometer, placing the cardboard cut-out leads to 
constructive and destructive interference at the outputs of the beam splitter (BS).\cite{bib116}}
    \label{fig:undetected}
\end{figure}

The first demonstration of quantum imaging using the principle of induced coherence~\cite{Zou1991PRL,Zou1991PRA} is in a method the investigators called imaging with undetected photons~\cite{Lemos2014,Kviatkovsky2020SciAdv,Viswanathan2021OE}. Here, signal-idler photon pairs are generated through SPDC from two nonlinear crystals that are pumped in succession. The idler photons from the first crystal pass through an object and are then sent through to the second crystal along with the pump. When the paths of the idler photons from the two crystals are aligned such that they are indistinguishable, coherence is induced between the signal photons from the two crystals, which are sent through separate paths to be recombined at a beam splitter, where their interference is viewed on a camera. When an object absorbs or scatters the idler photons from the first crystal, the idler photon becomes distinguishable from that produced in the second crystal, manifesting as a reduction in the interference visibility between the signal photons. One can thus infer the transmission or phase properties of the object through the signal photons alone, never detecting the idler photons which has interacted with the object. A setup for quantum imaging with undetected photons is shown in Fig.~\ref{fig:undetected} \cite{bib116}.

Interaction-free imaging with undetected photons has also since been demonstrated, allowing for the reduction of the required number of photons for imaging~\cite{Yang2022}.

The principle of induced coherence has also been used in the context of quantum microscopy ~\cite{Black2023Optica}. In this work, both the signal and idler photons from the first crystal probe the object. These photons are then passed through the “second crystal” which in fact is just the first crystal used in double pass. This procedure allows coherence to be induced between the signal photons and the idler photons of the two crystals. The phase structure is then extracted by a procedure analogous to that of classical phase-shifting holography.  This procedure demonstrated a doubling of the phase sensitivity and improving spatial resolution by a factor of 1.7.

\subsection{Multi-parameter quantum imaging:}
Entangled photon pairs generated from SPDC are hyperentangled in multiple degrees of freedom, such as time, frequency, position, and momentum. By simultaneously utilizing the correlation properties of entanglement in multiple parameters, it is possible to realize techniques such as snapshot hyperspectral imaging (SHI) \cite{Zhang2023} and light field imaging (LFI)\cite{Zhang2022,Zhang2024}. 

\begin{figure}
    \centering
    \includegraphics[width=1\linewidth]{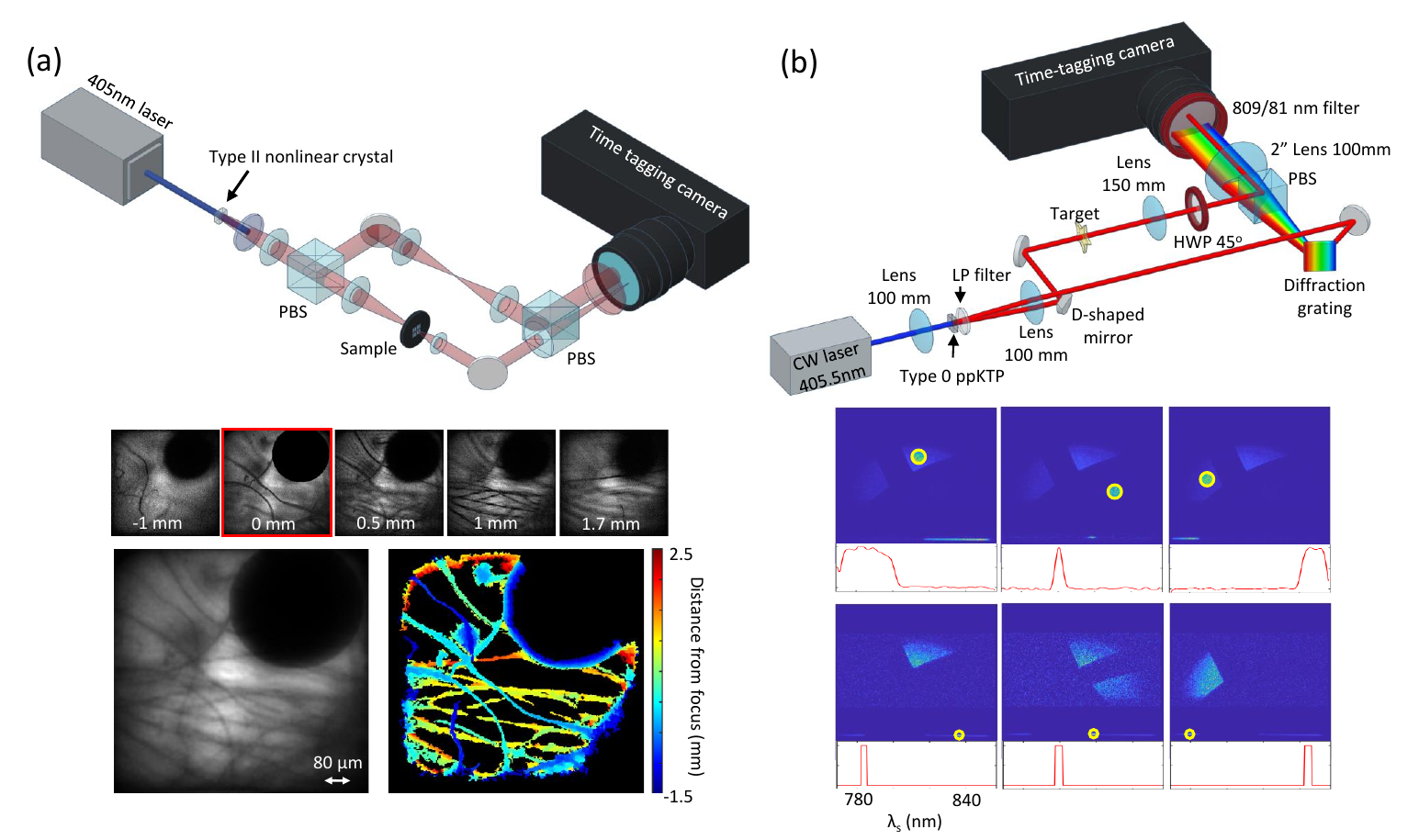}
    \caption{Multi-parameter quantum imaging, (a) Experimental setup of quantum correlation light field imaging (LFI) applied to microscopy, and the depth map image of fibers from a stack of lens cleaning tissue~\cite{Zhang2024}. (b) Experimental setup of quantum correlation snapshot hyperspectral imaging (SHI) with results demonstrating that the spectrum at any part of the image can be obtained or vice versa, where regions of an image with the selected wavelength can also be obtained~\cite{Zhang2023}.}
    \label{MPQI}
\end{figure}

Quantum SHI can be accomplished by measuring the position information through one photon of the pair (say the signal photon) and the spectral information measured through the partner photon (idler photon) ~\cite{Zhang2023}. Akin to simultaneously performing ghost imaging on the idler photon and ghost spectroscopy on the signal photon, position-spectral information of the photons can be simultaneously obtained. Similarly, quantum LFI can be accomplished by measuring the position information through one photon and the momentum/angular information measured through the partner photon to simultaneously capture the position-momentum information of the photons~\cite{Zhang2022,Zhang2024}. The setup and some results of quantum correlation SHI and quantum correlation LFI are shown in Fig.~\ref{MPQI}.

Classical methods for achieving SHI and LFI commonly require placing a microlens array over the imaging camera, with SHI also requiring an additional spectral filter array~\cite{Hagan2013,Yi2023,Wu2017}. This results in a reduction in the imaging resolution with a magnitude equal to the desired spectral or momentum resolution. Moreover, in conventional SHI, the use of a spectral filter array causes a large fraction of the photons to be discarded, proportional to the number of filters, making the technique highly photon-inefficient. For instance, the resolution of conventional LFI has the following constraint 
\begin{equation}
    N_{\text{tot}} = N_{\text{NF}} \times N_{\text{FF}},
\end{equation}
where $N_{\text{tot}}$ is the total number of camera pixels, $N_{\text{NF}}$ is the near field image resolution dictated by the number of mirolenses and $N_{\text{FF}}$ is the far field momentum resolution dictated by the number of pixels under each microlens. Typically one has $N_{{\text{NF}}} = 200\times200$ and $N_{{\text{FF}}} = 10\times10$ pixels for a $N_{{\text{tot}}} = 2000\times2000$ pixels camera. For SHI, there is a similar resolution constraint, but with the image resolution given by the number of pixels under each microlens and the spectral resolution given by the number of microlenses.

The quantum version of SHI and LFI does not have this resolution trade-off since each property can be measured on a separate camera using the full camera resolution. A much higher spectral or momentum resolution can thus be achieved without reducing the imaging resolution. Since no spectral filtering is needed in quantum SHI, the technique is also photon-efficient.

It was demonstrated for quantum SHI~\cite{Zhang2023} that at least a 2~nm spectral resolution over an 80~nm range can be obtained for each pixel of the image. For quantum LFI~\cite{Zhang2024}, a much higher angular resolution can be obtained compared to classical methods, and it was shown to have achieved up to an order of magnitude improvement in the depth of field compared to classical LFI. 

The idea of multi-parameter quantum imaging has also been extended to demonstrate interferometry-free phase microscopy~\cite{Hodgson2023, Zhang2024} whereby both the position and momentum information of the photons are simultaneously captured in a setup akin to that of quantum LFI with the phase profile of the sample extracted through the momentum information of the photons.

\section{Quantum-Inspired Imaging Techniques}
Here, we overview a few methods on ``quantum-inspired'' strategies that reproduce some of the advantages of quantum imaging techniques without requiring entanglement. These approaches are based on correlations, computational post-processing, and single-photon sensitive detectors to extend the reach of classical optics. Quantum-inspired methods thus occupy a middle ground: they do not use genuinely nonclassical resources, but they translate concepts originally developed in quantum optics into practical imaging schemes. Therefore, they can often be realized with simpler setups, higher photon flux, and reduced experimental overhead, while still achieving capabilities such as imaging under extremely low light conditions, background noise suppression, and noninvasive measurement. The following subsections are on two examples: classical ghost imaging, which mimics entangled-photon ghost imaging using thermal or computational light correlations, and imaging with single-photon sensitive cameras to capture ultraweak photon emission from biological systems.

\subsection{Classical Ghost Imaging}
Quantum GI utilizes the inherent correlations between spatially entangled photon pairs, this procedure has inspired the development of classical GI techniques which uses spatial correlations in the random speckle patterns of pseudo-thermal light source created by a laser beam passing through a rotating ground glass plate~\cite{bibz12,bibz16}. The pseudo-thermal light beam is then split into two, where one beam illuminates the object and is then collected by a bucket detector, which records the total intensity of the speckle pattern transmitted or reflected by the object \cite{Zerom2012ThermalGhostImaging}. The other beam is collected by a camera, which records the speckle pattern used to illuminate the object. This process would be repeated for thousands of random speckle patterns. The image of the object is then recovered by summing up the series of patterns, with the amplitude of each pattern given by its recorded intensity on the bucket detector. The number of patterns used would dictate the resolution of the image.  

This method has since been extended to computational GI~\cite{Shapiro2008,Katz2009} whereby the speckle patterns are now pre-programmed and generated through a beam modulation device such as a spatial light modulator or digital micromirror device. Here, the setup is further simplified in that only a bucket detector is needed to collect the intensity of the transmitted or reflected speckle patterns. A camera is no longer required as the patterns are pre-programmed and already known. 

\begin{figure}
    \centering
    \includegraphics[width=1\linewidth]{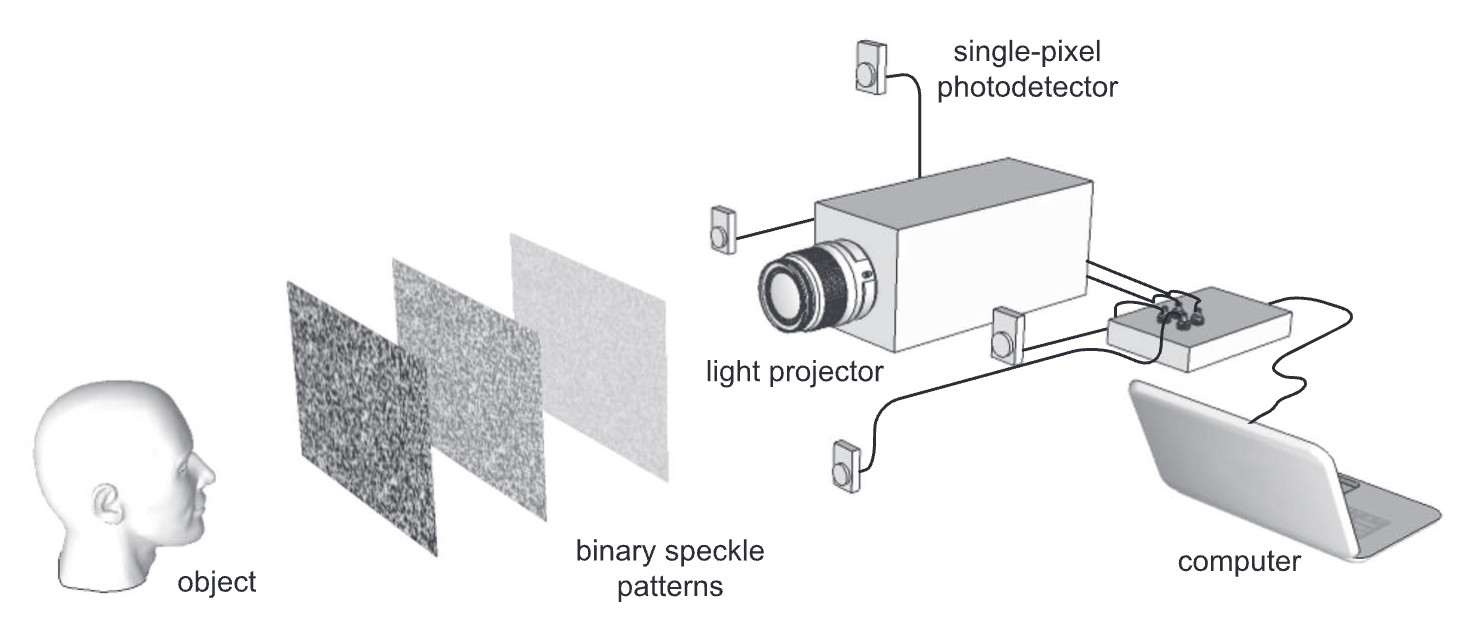}
    \caption{Computational 3D ghost imaging: the experimental setup for 3D surface reconstruction using a light projector to illuminate the object (a head) with computer-generated random binary speckle patterns. Four spatially separated single-pixel photodetectors capture the reflected light, and the signals recorded from these detectors are subsequently processed to reconstruct a computational image for each detection channel~\cite{Sun2013}.}
    \label{fig:CGI}
\end{figure}

Figure~\ref{fig:CGI} illustrates the use of computational GI for 3D imaging. A digital light projector generates computer-controlled random binary speckle patterns that illuminate an object, while four spatially separated single-pixel detectors capture the backscattered light. These intensity measurements, correlated with the known illumination patterns, yield four distinct two-dimensional reconstructions, each appearing under a virtual illumination direction tied to the detector location, which were used for 3D reconstruction of the object~\cite{Sun2013}.

Classical GI contains several advantages: it relaxes the need for entangled photon sources, can be implemented with broadband or incoherent light, and still supports imaging in scattering or noisy environments. Moreover, computational GI eliminates the need for a second physical beam, reducing complexity and allowing optimisation of the speckle pattern designs for faster convergence. Recent studies have shown the applicability of classical GI in biomedical imaging, where it indicates potential for noninvasive visualization of tissues under low-light conditions~\cite{Sun2013,Wang2016}. 

\subsection{Imaging with single-photon sensitive cameras}
The technique of direct imaging with quantum sensors, without needing energy source for illumination, can be categorized as classical imaging, since it does not use quantum entanglement or non-classical correlations typical of most quantum imaging techniques we discussed in the quantum imaging sections earlier. However, this may still be considered at the interface of quantum-inspired imaging because they use ultrasensitive detectors operating for the detection of extremely low intensity light at the single-photon level. Recent advances in low-noise and high-quantum-efficiency imaging systems, such as charge-coupling devices (CCD), electron-multiplying CCD (EMCCD), and scientific Complementary Metal-Oxide-Semiconductor (sCMOS) cameras, have significantly improved the spatial and temporal resolution of imaging single photon emissions from biological systems~\cite{Kobayashi2016}. CCD/EMCCD cameras have largely replaced less efficient detectors such as photomultiplier tubes (PMTs) in biological photon detection, which may open new avenues for biomedical imaging~\cite{kobayashi2000photon, kobayashi2009imaging, kobayashi2014highly}.

\begin{table}[t]
\renewcommand{\arraystretch}{1.3}
\centering
\caption{Representative benchmarks for ultraweak photon emission (UPE) imaging. Values are rough estimations and are setup-dependent.}
\label{tab:UPE}
\begin{tabular}{|c|c|c|}
\hline
\textbf{Metric} & \textbf{Typical values} & \textbf{Detection technology} \\
\hline\hline
Photon flux emitted from a biological tissue & $10^{2}$--$10^{3}$ photons/s/cm$^{2}$ & EMCCD, CCD, PMT \\
Dark count rates & $<10^{-3}$ e$^-$/pix/s & EMCCD /CCD/ sCMOS \\
Quantum efficiency (QE) & 80--95\% (500--700 nm) & EMCCD /CCD \\
Integration times & $>$10 min (imaging) & CCD/EMCCD \\
Spectral range & 300--900 nm & PMT (up to 600nm), CCD/EMCCD (up to 1000nm) \\
\hline
\end{tabular}
\end{table}

\begin{figure}
    \centering
    \includegraphics[width=1\linewidth]{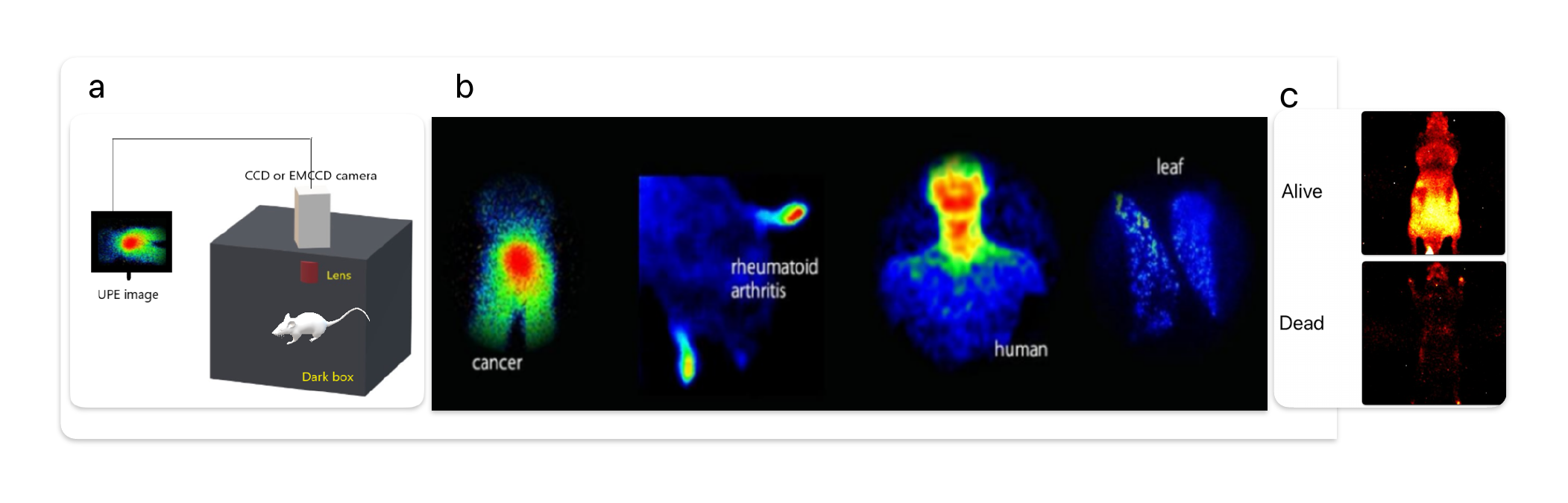}
    \caption{a) A simple UPE imaging setup with a quantum sensor, i.e. a CCD/EMCCD camera with high detection efficiency in the visible range, where the sample is placed in total darkness, e.g. in a dark enclosure with sufficient isolation from ambient light. b) Imaging UPE with a CCD camera from different living systems, from left to right, from a tumour in the body of a mouse, from the human body/face during sleep, and imaging abnormality in cut leaves~\cite{KobayashiLab}. c) UPE imaging of living and dead mouse body by EMCCD camera indicating UPE as the sign of vitality while their bodies kept at the same temperature \cite{Salari2025}.}
    \label{UPE1}
\end{figure}

Due to chemical reactions in cell metabolism, all living cells produce ultra-weak photon emission (UPE); however, the intensity of UPE is extremely weak ($\lesssim 10^3$ photons/sec/cm$^2$), making them hard to detect~\cite{CIFRA20142}. In humans, UPE is mostly due to metabolic oxidative stress processes that occur in the presence of reactive oxygen species. Over the last few decades, there has been significant progress in UPE research, including studies on how various diseases affect UPE~\cite{ives2014ultraweak}. UPE differs from black-body radiation and is a window into the fundamental processes of life, such as cellular metabolism and signalling~\cite{UPEUofC}. Moreover, UPE can serve as a diagnostic tool in biomedical applications, as variations in its intensity or spectral distribution may indicate the presence of specific pathological conditions, such as cancer~\cite{CancerUPE} or Alzheimer's disease~\cite{iScience}. Figure~\ref{UPE1}a shows a typical CCD-based UPE imaging setup, and Fig.~\ref{UPE1}b presents some examples of UPE images from different living samples, mice, humans, and plants. Table~\ref{tab:UPE} represent a benchmark for UPE imaging.

CCD and EMCCD cameras often use long integration times, highly sensitive optics, and spectral filters to capture very weak light to form a meaningful UPE image, and moreover, allowing us to track spatial changes in UPE patterns over time in a dynamic manner, where capturing images in shorter time intervals and consequently making a video of variations with sequences of those captured images. More advanced approaches, such as time-resolved and hyperspectral UPE imaging, are now being developed. These methods measure the total number of emitted photons and reveal their wavelength distribution and how the light intensity may decay over time. Such information can provide more information about underlying metabolic processes and the role of oxidative stress in biological systems~\cite{Cifra2015}. Beyond diagnostics, UPE imaging can be used in neuroscience, plant physiology, and vitality~\cite{Salari2025}, where it can image and map neuronal activity~\cite{Wang2019}, track abnormalities in different living systems, and reveal plant stress.

\section{Summary and Conclusion}

\begin{table}[h]
\renewcommand{\arraystretch}{1.7}
\centering
\rowcolors{1}{green!20}{green!20}

\caption{General comparison between conventional and quantum imaging methods.}
\label{tab:Comparison1}
\begin{tabular}{p{3.5cm} p{7cm} p{7cm}}
\hline
\textbf{Feature / Aspect} & \textbf{Conventional Imaging} & \textbf{Quantum Imaging} \\
\hline\hline
Light Source & Uses classical light sources (X-rays, ultrasound, visible/IR lasers, RF waves, etc.) & Uses nonclassical light sources (entangled photons, squeezed states, N00N states) \\

Information Carriers & Independent photons without quantum correlations & Photons with quantum correlations (entanglement, squeezing, hyperentanglement) \\

Noise Performance & Limited by shot noise; high illumination intensity needed for better SNR & Can achieve sub shot-noise imaging; tolerant to high background noise \\

Radiation Dose & Often requires higher photon dose for sufficient image quality (possible photodamage) & Potentially much lower photon dose due to higher sensitivity \\

Real-time Capability & Established for many methods (ultrasound, fluoroscopy) & Often limited by current detector speed and photon flux \\

Technological Maturity & Fully mature, integrated into clinical workflows & Mostly in experimental/research stage, with few pre-clinical demonstrations \\
\hline
\end{tabular}
\end{table}

\begin{table}[t]
\renewcommand{\arraystretch}{1.3}
\centering
\rowcolors{1}{blue!20}{blue!20}
\caption{Numerical benchmarks comparing conventional and quantum imaging. Values are representative; performance depends on setup and sample. Technology Readiness Level (TRL) indicates how mature a technology is, from early concept (TRL 1) to proven operational system (TRL 9) \cite{TRL}. For optical techniques here, TRL is typically estimated by matching the current stage of development to the TRL definitions: e.g. proof-of-principle lab experiment (TRL 3–4), prototype demonstrated in a relevant environment (TRL 5–6), or field-tested/clinically validated system (TRL 7–8), with commercial deployment corresponding to TRL 9.}
\label{tab:Comarison3}
\begin{tabular}{ccccc}
\hline
\textbf{Modality} & \textbf{Axial res. ($\mu$m)} & \textbf{Depth (mm)} & \textbf{Flux (photons/pairs/s)} & \textbf{TRL} \\
\hline\hline
OCT (classical) & 3--10 & 1--3 & $10^{13}$--$10^{15}$ & 9 \\
QOCT & $\sim$2--5$^\dagger$ & $\lesssim 1$ & $10^{5}$--$10^{7}$ & 3--4 \\
\hline
Confocal / Widefield Microscopy & Diffraction-limited & sub-mm & High (mW--W laser) & 9 \\
Quantum Microscopy & Sub-SNL phase & superficial layers & $10^{5}$--$10^{7}$ pairs/s & 2--4 \\
\hline
Classical Ghost Imaging & Pattern-limited & scene dep. & $10^{10}$--$10^{14}$ & 6--8 \\
Quantum Ghost Imaging & Correlation-limited & scene dep. & $10^{5}$--$10^{7}$ & 2--4 \\
\hline
SHI / LFI (classical) & Resolution trade-offs & scene dep. & High (mW) & 7--8 \\
Quantum SHI / LFI & Full res. retained & scene dep. & $10^{7}$ & 3--4 \\
\hline
UPE Imaging & N/A & surface & $10^{2}$--$10^{3}$/cm$^2$ & 5--7 \\
\hline
\end{tabular}

\vspace{4pt}
\footnotesize
$^\dagger$QOCT may achieve $\sim$2x axial resolution vs. OCT under ideal entangled-photon conditions; limited in practice by photon flux and detection speed.
\end{table}

\begin{table}[h]
\renewcommand{\arraystretch}{1.7}
\centering
\rowcolors{1}{green!20}{green!20}
\caption{Specific biomedical advantages of quantum imaging methods compared to conventional counterparts.}
\label{tab:Comparison2}
\begin{tabular}{p{4cm} p{4cm} p{9.5cm}}
\hline
\textbf{Conventional Modality} & \textbf{Quantum Counterpart} & \textbf{Biomedical Advantages of Quantum Version} \\
\hline
\multirow[t]{2}{4cm}{Optical Coherence Tomography (OCT)} & \multirow[t]{2}{4cm}{Quantum OCT (QOCT)} & Potential advantages include enhanced axial resolution and even-order dispersion cancellation; SNR, penetration depth, and scattering robustness depend on flux, bandwidth, and detection efficiency, and can be limited by today’s sources and detectors. \\

\multirow[t]{2}{4cm}{Classical Microscopy} & \multirow[t]{2}{4cm}{Quantum Optical Microscopy} & Achieves the same image quality with fewer probe photons (lower light dose) without altering single-photon absorption (not two-photon excitation); better for imaging light-sensitive biological samples. \\

\multirow[t]{2}{4cm}{Classical Ghost Imaging} & \multirow[t]{2}{4cm}{Quantum Ghost Imaging (QGI)} & Higher contrast and visibility; no featureless background; higher SNR at low light levels \\

\multirow[t]{2}{4cm}{Conventional Hyperspectral or Light Field Imaging} & \multirow[t]{3}{4cm}{Multi-parameter Quantum Imaging (Quantum SHI / Quantum LFI)} & Simultaneous spatial-spectral or spatial-momentum imaging without resolution loss; high photon efficiency (no filter array losses); improved depth of field and spectral resolution. \\

\multirow[t]{2}{4cm}{Conventional Low-Light Imaging} & \multirow[t]{3}{4cm}{Ultraweak Photon Emission (UPE) Imaging with Quantum Sensors} & High-sensitivity CCD/EMCCD cameras for potential non-invasive biomarker detection; avoids external illumination. These cameras may also be used for classical imaging at safe light levels, otherwise it may harm their sensor.\\
\hline
\end{tabular}
\end{table}

\begin{table}[h!]
\renewcommand{\arraystretch}{1.7}
\centering
\rowcolors{1}{gray!20}{gray!20}
\caption{Key challenges of quantum imaging methods for biomedical applications and potential solutions.}
\label{tab:Challenges}
\begin{tabular}{p{4cm}p{6.5cm}p{7cm}}
\hline
\textbf{Challenge} & \textbf{Description / Impact} & \textbf{Potential Future Solution} \\
\hline
\multirow[t]{2}{4cm}{Entanglement degradation and limited penetration depth} & Quantum entanglement and correlations are easily degraded through scattering, losses and mode mixing. & Development of noise-resilient quantum states (e.g., decoherence-free subspaces, error correction codes) and improved shielding or cryogenic stabilization. Also, hybrid approaches combining quantum imaging with classical modalities (e.g., ultrasound or MRI guidance), adaptive optics, and novel near-infrared entangled sources may mitigate limited penetration depth issue. \\

\multirow[t]{2}{4cm}{Entanglement/correlation quality of quantum source} & The achievable imaging resolution is often dictated by how good is the entanglement/correlation of the photon source. These are often not good enough to allow diffraction limited microscopy even without entanglement degradation. & Using brighter, narrower-bandwidth SPDC sources with better mode matching and active stabilization (to maintain high-quality entanglement/correlations) and by applying quantum error correction or entanglement distillation techniques\\

Low imaging speed & Slow acquisition times hinder real-time imaging needed in biomedical contexts. & Advances in fast single-photon detectors, parallelized acquisition methods, and integrated photonic circuits for faster processing. \\

\multirow[t]{2}{4cm}{Lack of ideal detector/camera} & The high efficiency ones either has low spatial resolution (e.g. SNSPD, $>90\%$ QE) or slow speed (e.g. CCD/EMCCD cameras). The high speed ones has low efficiency (~10\%) (e.g. SPAD array and Timepix cameras).& Next generation cameras (e.g. SNSPD array cameras) should solve the problem\\

Low photon number regime & Often used to avoid sample damage, but can limit achievable image quality. & Improved quantum light sources with tunable brightness and wavelength, along with non-linear quantum amplifiers \\
\multirow[t]{2}{4cm}{Calibration and standardization} & Lack of standardized protocols and calibration methods limits reproducibility and clinical translation. & Establishment of international standards, reference phantoms for quantum imaging, and robust calibration protocols validated in multi-center studies. \\

\hline
\end{tabular}
\end{table}

In this review, we have presented an overview of quantum optical imaging methods and indicated the potential of quantum methods that may lead to the discovery of new quantum protocols for real-life applications in biology and medicine. Quantum imaging techniques have the potential to provide higher resolution, improved contrast in scattering tissue, and functional imaging at ultra-low light levels, thereby minimizing photodamage and reducing the risk of damaging sensitive biological samples. These features are particularly attractive for neuroscience, ophthalmology, and regenerative medicine, where non-invasive, high-sensitivity imaging modalities are essential. 

Tables~\ref{tab:Comparison1} and ~\ref{tab:Comarison3} give an overview of differences between conventional and quantum imaging techniques qualitatively and quantitatively, and Table~\ref{tab:Comparison2} summarizes the advantages of quantum techniques over their conventional counterparts.

Despite several advantages, quantum imaging methods for biomedical applications face a range of significant challenges. Environmental noise and decoherence can degrade quantum information and entanglement, reducing imaging performance. The low signal-to-noise ratio (SNR) is particularly problematic due to the sensitivity of quantum systems to loss and absorption in biological tissues. This challenge is exacerbated by the limited penetration depth of quantum techniques (due to loss and using the wavelengths that penetrate less in the tissues), which struggle to image deep tissue structures effectively. In addition, the achievable spatial resolution is strongly constrained by the entanglement/correlation quality of the photon source itself; current sources often do not provide sufficiently high-quality correlations to reach diffraction-limited microscopy. The low acquisition speed of current quantum technologies further hampers their practical use, as real-time imaging is often required in biomedical settings. The low photon-number regime often employed in quantum imaging, also presents difficulties in terms of achieving sufficient image quality. Detector limitations compound these issues, as high-efficiency detectors typically sacrifice either spatial resolution or speed, while high-speed detectors usually have low quantum efficiency; however, a new generation of cameras may help solve this issue (e.g. superconducting nano-wire single photon detectors (SNSPDs) cameras~\cite{NatureSNSPD}). Moreover, the complexity of data produced by quantum methods may necessitate advanced computational techniques for accurate image reconstruction and may complicate integration with existing biomedical workflows. Finally, the absence of standardized calibration protocols and reference phantoms restricts reproducibility and slows clinical translation, underscoring the need for international standards and multi-center validation.

Table~\ref{tab:Challenges} is a summary of key challenges of quantum imaging methods for biomedical applications with some potential future solutions to mitigate those challenges to make quantum imaging applicable in real-life. Advances in integrated quantum photonics and noise-resilient quantum states are expected to mitigate decoherence and improve robustness in biological environments. The combination of quantum-enhanced light sources with machine-learning–assisted image reconstruction may both boost signal-to-noise ratios and accelerate data processing, bringing real-time imaging closer to feasibility. Hybrid approaches that merge quantum techniques with established modalities such as MRI, ultrasound, or adaptive optics could extend penetration depth and broaden clinical applicability. 

Together, these efforts suggest that the path forward will likely rely on a multi-disciplinary integration of quantum optics, biomedical engineering, and regulatory science to transform current limitations into opportunities for innovation.
With these advances, quantum imaging may evolve from proof-of-principle demonstrations into a transformative modality that complements and, in certain niches, surpasses the capabilities of conventional biomedical imaging technologies.

\paragraph{Acknowledgments}
VS, YZ, DP, RB, EK, CS, and DO acknowledge funding support from the Natural Sciences and Engineering Research Council of Canada (NSERC) through the Quantum Alliance Grant on Quantum Enhanced Sensing and Imaging (QuEnSI), and VS, DE, CS, and DO acknowledge the funding support from the National Research Council of Canada through its Quantum Sensors Challenge Program.

\bibliography{QI-v1}

\end{document}